\documentclass[aps,prd,superscriptaddress,amsmath,amssymb,groupedaddress]{revtex4}
\usepackage{amsmath}
\usepackage{graphicx}
\usepackage{latexsym}
\usepackage{bm}
\usepackage{array}
\usepackage{dcolumn}
\usepackage{longtable}
\usepackage{tabularx}
\usepackage[usenames,dvipsnames]{color}
\newcommand{\be}{\begin{equation}}\newcommand{\ee}{\end{equation}}
\newcommand{\bea}{\begin{eqnarray}}\newcommand{\eea}{\end{eqnarray}}
\newcommand{\brr}{\begin{array}}\newcommand{\err}{\end{array}}
\newcommand{\bit}{\begin{itemize}}\newcommand{\eit}{\end{itemize}}
\newcommand{\ben}{\begin{enumerate}}\newcommand{\een}{\end{enumerate}}

\newcommand{\bbm}{\begin{bmatrix}}\newcommand{\ebm}{\end{bmatrix}}
\newcommand{\ba}{\begin{array}}
\newcommand{\ea}{\end{array}}
\newcommand{\G}{\textbf}

\newtheorem{mydef}{Definition}
\newtheorem{Lemma}{Lemma}
\newtheorem{theorem}{Theorem}
\newcommand{\bd}{\begin{mydef}} \newcommand{\ed}{\end{mydef}}
\newcommand{\bthe}{\begin{theorem}} \newcommand{\ethe}{\end{theorem}}
\newcommand{\ble}{\begin{Lemma}} \newcommand{\ele}{\end{Lemma}}

\newcommand{\dr}{\mathrm{d}}
\newcommand{\mpsi}{\boldsymbol{\psi}}

\def\ha{\frac{1}{2}}

\def\intx{\int \! \! \mathrm{d}^3 \textbf{x}}
\def\intk{\int \! \! \mathrm{d}^3 \textbf{k}}

\def\ph{\varphi}
\def\lan{\langle}
\def\lf{\left}

\def\non{\nonumber}\def\pa{\partial}\def\ran{\rangle}

\def\ri{\right}
\def\al{\alpha}\def\bt{\beta}\def\ga{\gamma}
\def\de{\delta}\def\ep{\epsilon}
\def\te{\vartheta}\def\te{\Theta}
\def\si{\sigma}
\def\om{\omega}

\def\1{{_{1}}}\def\2{{_{2}}}

\newcommand{\ide}{1\hspace{-1mm}{\rm I}}

\def\noHe0{:\;\!\!\;\!\!:H_e(0):\;\!\!\;\!\!:}
\def\noHm0{:\;\!\!\;\!\!:H_\mu(0):\;\!\!\;\!\!:}

\def\boldsymbol#1{{\bm #1}}

\def\lan{\langle}
\def\lf{\left}

\def\non{\nonumber}
\def\pa{\partial}\def\ran{\rangle}

\def\ri{\right}

\def\al{\alpha}\def\bt{\beta}\def\ga{\gamma}
\def\de{\delta}
\def\ep{\epsilon}\def\te{\theta}
\def\te{\Theta}
\def\si{\sigma}
\def\om{\omega}

\def\1{{_{1}}}\def\2{{_{2}}}
\def\I{{_{\rm{I}}}}\def\II{{_{\rm{II}}}}

\definecolor{darkred}{rgb}{.8,0,0}

\definecolor{darkblue}{rgb}{0,0,.7}

\begin{document}

\title{Dynamical generation of field mixing via flavor vacuum condensate}

\author{M.~Blasone}
\email{blasone@sa.infn.it}
\affiliation{Dipartimento di Fisica, Universit\`a di Salerno, Via Giovanni Paolo II, 132 84084 Fisciano, Italy \& INFN Sezione di Napoli, Gruppo collegato di Salerno, Italy}
\author{P.~Jizba}
\email{p.jizba@fjfi.cvut.cz}
\affiliation{FNSPE, Czech Technical
University in Prague, B\v{r}ehov\'{a} 7, 115 19 Praha 1, Czech Republic\\}
\affiliation{ITP, Freie Universit\"{a}t Berlin, Arnimallee 14,
D-14195 Berlin, Germany}
\author{N.E.~Mavromatos}
\email{Nikolaos.Mavromatos@kcl.ac.uk}

\affiliation{Theoretical Particle Physics and Cosmology Group, Department of Physics, King's College London, Strand WC2R 2LS, UK}

\author{L.~Smaldone}
\email{lsmaldone@sa.infn.it}
\affiliation{Dipartimento di Fisica, Universit\`a di Salerno, Via Giovanni Paolo II, 132 84084 Fisciano, Italy \& INFN Sezione di Napoli, Gruppo collegato di Salerno, Italy}

\begin{abstract}
In this paper we study dynamical chiral symmetry breaking of a generic quantum field theoretical
model with global $SU(2)_L \times SU(2)_R \times U(1)_V$ symmetry.
By purely algebraic means we analyze the vacuum structure for different symmetry breaking
schemes and show explicitly how  the  ensuing non-trivial flavor vacuum condensate, originally introduced in connection with neutrino oscillations, characterizes the dynamical generation of field mixing. In addition,
with the help of Ward--Takahashi identities we demonstrate the emergence
of the correct number of Nambu--Goldstone modes in the physical spectrum.
\end{abstract}

\vspace{-1mm}

\maketitle

\section{Introduction}

Particle mixing represents one of the most exciting topics in modern theoretical and experimental physics~\cite{partref}.
The mechanism of quark mixing was theoretically first described by Cabibbo~\cite{cabbibo} in 1960's and by Kobayashi and Maskawa~\cite{maskawa} in the early 1970's. Neutrino mixing was formulated, in parallel, by Pontecorvo~\cite{Pontecorvo} and by Maki, Nakagawa, and Sakata~\cite{sakata}.

Particle mixing can be treated from various standpoints. The most common description is in the framework of quantum mechanics~\cite{Pontecorvo,partref}. More modern approaches to mixing involve both perturbative \cite{NeutPheno} and non-perturbative quantum field theoretical (QFT) methods \cite{Mixing,BJVa}. The basic features of a non-perturbative treatment of flavor mixing in the QFT framework were first exploited in Ref.~\cite{BlaVit95} and consequently discussed in many subsequent papers (see e.g. Ref.~\cite{Mixing}). There, it was observed that field mixing is not the same as wave-function (i.e., first-quantized) mixing and, in fact, corrections to neutrino oscillation formula were found in the QFT setting~\cite{BHV99,ChargesBJV}.
The origin of these discrepancies can be retraced to the non-trivial nature of the mixing transformation~\cite{BlaMavVit},
which implies a rich structure of the \textit{flavor vacuum}, characterized by a condensate of fermion-antifermion pairs~\cite{BlaVit95}. A similar, though not identical, conclusions were also formerly reached in the context of grand unified theories~\cite{Chang}.

Furthermore, in Ref.~\cite{mavroman}, it was proved that the aforesaid flavor vacuum structure naturally arises when mixing is dynamically generated by an effective four-fermion interaction, within a string inspired scenario. In that context four-fermion interaction appears as a consequence of string scattering with point-like brane defects in the structure of space-time. Since the flavor is not preserved during such a scattering, the outgoing states exhibit a flavor mixing. In Ref.~\cite{DynMix}, where the Nambu--Jona Lasinio model~\cite{NJL} was studied, similar results were also found.

In this paper, we demonstrate that an analogous vacuum structure inevitably appears when mixing is dynamically generated,
from an algebraic (and hence manifestly non-perturbative) point of view.
In fact, the only requirement we enforce in our considerations is the invariance of the Lagrangian under the global $SU(2)_L \times SU(2)_R \times U(1)_V$ symmetry. In particular, we do not explicitly consider axial $U(1)_A$ symmetry, which may be  broken from the very beginning~\cite{tHooft1976}. Our aim here is to start with the aforesaid global symmetry and analyze symmetry-breaking schemes that are pertinent for the dynamical generation of masses and, in particular, field mixing.
We envisage that a particular breaking scheme is realized via spontaneous
symmetry breaking (SSB) phase transition whose pattern is dictated by the
potential-energy term in the original or effective-action Lagrangian. In fact,
the specific form of the potential is immaterial for our description. Apart from
discussing explicit structural forms of broken-phase vacua we also show how the correct number of Nambu--Goldstone (NG) bosons appear in the
physical spectrum of the full theory. It turns out that the presence of fermion-antifermion pairs with different masses is a \textit{necessary} condition, in order to dynamically generate mixing in the present context.
The novelty of the presented approach is the emergence, after symmetry breaking, of flavor vacuum and ensuing Hilbert space~\cite{BlaVit95}, which is unitarily inequivalent to the usually adopted Hilbert space for  mass eigenstates.
It should be stressed that such unitary inequivalence is not a mere mathematical feature, being at the origin of phenomenological effects, e.g. corrections to the usual neutrino oscillation formulas~\cite{BHV99}.

In contrast to most of the works on dynamical generation of masses and mixing~\cite{QCDmassmix,DynMassMix},
this paper is primarily devoted to a study of general algebraic aspects of field mixing and vacuum structure
rather than to phenomenological issues. This approach allows us to identify the flavor vacuum condensate structure, originally obtained in the study of the generator of mixing transformations~\cite{BlaVit95}.

The paper is organized as follows: in Section~\ref{ChAl} we review all the ways in which the chiral-charge conservation can  explicitly be broken
when a generic mass-matrix term is added to a chirally symmetric action.
In Section~\ref{LBcon} we formulate the Ward--Takahashi (WT) identities for three quintessential chiral SSB schemes. This is done with the help of the so-called Umezawa's $\varepsilon$-term prescription~\cite{Umezawa,FujPap}. In doing so we demonstrate that in order to generate mixing dynamically: a) the residual $U(1)_V\times U(1)^\boldsymbol{m}_V$ symmetry, encountered on the classical level, must be dynamically broken by flavor vacuum condensate, b) in this respect, SSB must happen in two sequential steps. In Section~\ref{meanmix}, we use the mean field approximation (MFA) to prove how the aforementioned vacua are algebraically related to masses and mixing generators and we recover flavor vacuum from Refs.~\cite{BlaVit95,Mixing,BlaMavVit,mavroman}. Various remarks and generalizations are addressed in
the concluding Section~\ref{conclusioni}. For the reader's convenience, the paper is supplemented with three appendices. In Appendix~\ref{wt} we review the
proof of WT identities with $\varepsilon$-term prescription, while in Appendices~\ref{ssbu1u1} and~\ref{tab}
we discuss several technical issues related to Section~\ref{LBcon}.
\section{Explicit chiral symmetry breaking and fermion mixing}\label{ChAl}
%
Let us consider  the Lagrangian density $\mathcal{L}$ that is
invariant under the global \textit{chiral-flavor} group $G = SU(2)_L \times SU(2)_R \times U(1)_V $. Let the fermion field
be a flavor doublet
\be
\boldsymbol{\psi} \ = \
\bbm
\tilde{\psi}_1 \\ \tilde{\psi}_2
\ebm \, .
\ee
Under a generic chiral-group transformation $\G g$, the field $\boldsymbol{\psi}$ transforms to $\boldsymbol{\psi}'$ where~\cite{Miransky}
\bea
 \boldsymbol{\psi}'  =  \G g \mpsi  \ = \ \exp\lf[i \lf(\phi+\boldsymbol{\om} \cdot \frac{\boldsymbol{\si}}{2}+\boldsymbol{\om}_5 \cdot\frac{\boldsymbol{\si}}{2} \ga_5\ri)\ri] \mpsi \, . \label{chigr}
\eea
Here $\si_j \, , j=1,2,3$ are the Pauli matrices and $\phi, \boldsymbol{\om}$,  $\boldsymbol{\om}_5$ are real-valued
transformation parameters of $G$. Noether's theorem implies the conserved vector and axial currents
\bea
J^\mu & = & \overline{\boldsymbol{\psi}}\ga^\mu \boldsymbol{\psi} \, , \\[2mm]
\boldsymbol{J}^{\mu} & = & \overline{\boldsymbol{\psi}}\gamma^\mu \frac{\boldsymbol{\si}}{2} \boldsymbol{\psi}\, ,\\[2mm]
\boldsymbol{J}^\mu_{5} & = & \overline{\boldsymbol{\psi}}\gamma^\mu \gamma_5 \frac{\boldsymbol{\si}}{2} \boldsymbol{\psi} \, ,
\eea
and the ensuing conserved charges
\bea
Q & = & {\intx} \, \boldsymbol{\psi}^\dagger \boldsymbol{\psi} \, , \\[2mm]
\boldsymbol{Q} & = & \intx \, \mpsi^\dagger  \frac{\boldsymbol{\si}}{2}  \mpsi \, ,\\[2mm]
\boldsymbol{Q}_5 & = & \intx \, \mpsi^\dagger \frac{\boldsymbol{\si}}{2}  \ga_5 \mpsi  \, .
\eea
From these we recover the Lie algebra of the chiral-flavor group $G$, i.e.
\bea \label{su23}
&& \label{su21} \lf[Q_i,Q_j \ri] \ = \ i \,\varepsilon_{ijk} Q_k \, , \quad
 \lf[Q_i, Q_{5,j} \ri] \ = \ i \, \varepsilon_{ijk} Q_{5,k} \, , \quad
  \lf[Q_{5,i}, Q_{5,j} \ri] \ = \ i \,\varepsilon_{ijk}{Q_{k}} \, ,\quad \lf[Q , { Q_{5,j}} \ri] \ = \ { \lf[Q , Q_{j} \ri]}  \ = \ 0 \, .
\eea
Here $i,j,k=1,2,3$ and $\varepsilon_{ijk}$ is the Levi-Civita pseudo-tensor.

For massless fermions the Lagrangian
is invariant under both the
flavor and axial flavor transformations.
The chiral symmetry is explicitly broken when a mass term
\be
\mathcal{L}_M \ = \  - \overline{\boldsymbol{\psi}}\, \mathbb{M} \, \boldsymbol{\psi} \, ,
\ee
is added to $\mathcal{L}$. In fact, one can easily verify, that~\cite{Miransky}
\bea
\pa_\mu J^\mu & = & 0 \, , \quad \pa_\mu \boldsymbol{J}^{\mu} \ = \  \frac{i}{2} \overline{\boldsymbol{\psi}} \, \lf[\mathbb{M}, \boldsymbol{\si} \ri] \, \boldsymbol{\psi} \, , \quad \pa_\mu \boldsymbol{J}^\mu_{5} \ = \  \frac{i}{2} \overline{\boldsymbol{\psi}} \gamma_5 \lf\{\mathbb{M}, \boldsymbol{\si}\ri\} \boldsymbol{\psi} \, .
\label{consveccurax}
\eea
Note, that these relations do not presuppose any specific form of the original  chirally symmetric action.
We now demonstrate how particular choices of the mass matrix $M$ can affect the structure of the residual (unbroken) subgroup.\\

{\bf i)} Let $\mathbb{M} =  m_0 \ide$ where $\ide$ is the identity matrix, then~\eqref{consveccurax} reduces to
\bea
\pa_\mu J^\mu \ = \ \pa_\mu \boldsymbol{J}^{\mu} \ =  \ 0  \, , \;\;\;\;\;
\pa_\mu \boldsymbol{J}^\mu_{5} \ = \ {i \, m_0 \, \overline{\boldsymbol{\psi}} \, \gamma_5\, \boldsymbol{\si} \, \boldsymbol{\psi} \, ,}
\eea
i.e., the scalar and vector currents remain conserved and the broken-phase symmetry is $H = U(2)_V$.\\

{\bf ii)} If $\mathbb{M}   = m_0   \ide  +  m_3  \, \si_3$, then Noether currents satisfy
\bea
\pa_\mu J^{\mu} & = & \pa_\mu J_3^{\mu} \ = \  0  \, , \;\;\;\;\;
\pa_\mu J_1^{\mu} \ = \ { -} m_3 \,  \overline{\boldsymbol{\psi}} \, \si_2\, \boldsymbol{\psi} \, ,
\;\;\;\;\;\pa_\mu J_2^{\mu}  =   {m_3 \, \overline{\boldsymbol{\psi}} \, \si_1  \, \boldsymbol{\psi}} \, , \;\;\;\;\; \pa_\mu \boldsymbol{J}^\mu_{5} \ \neq \ {0}\, ,
\eea
thus reducing the residual symmetry to $H = U(1)_V \times U(1)^3_V$ where $U(1)^3_V$ represents transformations generated by $Q_3$.\\

{\bf iii)} Finally we consider $\mathbb{M}   =  m_0   \ide  +  m_3   \si_3  +  m_1  \si_1 + m_2 \si_2$. In this case Eq.~\eqref{consveccurax} yields
\bea
\pa_\mu J^{\mu}  & = &  0  \, , \;\;\; \pa_\mu \boldsymbol{J}^\mu_{5}  \neq  {0}\, , \label{2.15.aa}\\[2mm] \label{16}
\pa_\mu J_1^{\mu} & = &  \overline{\boldsymbol{\psi}} \, \lf(m_2\,\si_3-m_3\,\si_2\ri)\, \boldsymbol{\psi} \, , \\[2mm]
\pa_\mu J_2^{\mu} & =  & -\overline{\boldsymbol{\psi}} \,\lf(m_1 \si_3 \ { -} \ m_3\si_1 \ri) \, \boldsymbol{\psi} \, ,  \\[2mm]
\pa_\mu J_3^{\mu} & = & \overline{\boldsymbol{\psi}} \, \lf(m_1\si_2-m_2 \si_1 \ri)\, \boldsymbol{\psi} \, , \;\;\;\;\; \, . \label{18}
\eea
It might seem that the residual symmetry is then  $H = U(1)_V$ which is associated
with the charge $Q$. R\^{o}le of this symmetry can be understood by considering the
\textit{flavor-charges}~\cite{ChargesBJV}
\bea
&&Q_\I \ \equiv \ \ha Q \, + \, Q_3 \, ,\;\;\;\;\;\;\;
Q_\II \ \equiv \  \ha Q \, - \, Q_3 \, ,
\label{flav2}
\eea
where the total flavor-charge is $Q = Q_\I  +  Q_\II$.  However, let us observe that from Eqs.\eqref{16}-\eqref{18}
there is yet another conserved current, namely
\be
J^\mu_\boldsymbol{m} \ \equiv \ \sum^3_{k=1} \, m_k \, J^\mu_k \, ,
\ee
and an ensuing conserved charge $Q_{\boldsymbol{m}}=\sum^3_{k=1}m_k Q_k$, where $\boldsymbol{m}=(m_1,m_2,m_3)$. For a future convenience we denote this residual symmetry as $U(1)_V \times U(1)^\boldsymbol{m}_V$. Actually this is isomorphic to $U(1)_V \times U(1)^3_V$, encountered in the case ii). However, in the following sections we will see that the classical Noether's charge $Q_\boldsymbol{m}$ is dynamically broken by the flavor vacuum condensate, while the charge $Q$ will still remain conserved. Hence the only residual symmetry at QFT level will be $U(1)_V$. This is indeed compatible with the experimental observation that, for neutrino oscillation, only a single charge (total flavor) is conserved \cite{BilPet}. In the case of neutrinos \cite{ChargesBJV,BJSun}, a charged current term is present in the Lagrangian, which is not invariant under the transformation generated by $Q^\boldsymbol{m}$.
\section{Dynamical generation of masses and mixing  in the flavor vacuum framework}\label{LBcon}

In the previous section we reviewed how the mass generation and mixing phenomena are characterized
by the residual symmetry. The subsequent considerations will be done in a full QFT framework, employing the flavor vacuum approach to mixing~\cite{BlaVit95}.

To proceed, let us recall~\cite{Miransky,TK} that the key signature of SSB is the existence of some
local operator(s) $\phi(x)$ so that on the vacuum $|\Omega\ran$.
\be \label{cssb}
\lan \lf[N_i, \phi(0) \ri] \ran \ = \  \lan \ph_i(0)\ran \  \ \equiv \ v_i \  \neq \ 0 \, ,
\ee
where $\lan \ldots \ran \equiv \lan \Omega | \ldots| \Omega \ran$. Here $v_i$ are the \textit{order parameters} and $N_i$ represent group generators from the quotient space $G/H$. In our case $N_i$ will
be given by $\boldsymbol{Q}$ and ${\boldsymbol{Q}}_5$ according to the chosen SSB scheme.

By analogy with quark condensation in QCD~\cite{Miransky,QCDmassmix}, we will limit our considerations to  order parameters that are condensates of fermion-antifermion pairs.
To this end we introduce the following composite operators
\bea
\Phi_k \ = \  \overline{\boldsymbol{\psi}}\, \si_k \,  \mpsi \, , \quad  \Phi^5_k \ = \  \overline{\boldsymbol{\psi}}\, \si_k \, \ga_5 \mpsi\, , \quad  \sigma_0 \ \equiv\ \ide\, ,   \label{bil35}
\eea
with $k = 0,1,2,3$. For simplicity we now assume $\lan \boldsymbol{\Phi}^5\ran =0$. This hypothesis does not affect our reasoning.

Let us now look in some detail at three SSB schemes $G \rightarrow H$ outlined in Section~\ref{ChAl}. \\

{\bf i)} SSB sequence corresponding to a single mass generation is~\cite{Miransky, FujPap}
\be
SU(2)_L \times SU(2)_R \times U(1)_V \ \! \longrightarrow \ \! SU(2)_V \times U(1)_V \, .
\label{SSB-scheme-1}
\ee
Note that $H \sim U(2)_V $. The broken-phase symmetry (which corresponds to dynamically generated mass matrix $M  =  m_0 \ide$)
is characterized by the order parameter
\bea
\lan \Phi_0 \ran \ = \ v_0  \ \neq 0 \, , \qquad \lan \Phi_k \ran \ = \ 0 \, , \quad  k = 1,2,3 \, . \label{diag1}
\eea
One can easily check that this is invariant under the residual group $H$ but not under the full chiral group $G$.

In order to discuss the NG modes it is convenient employ the WT identity~\eqref{diag22} within the $\varepsilon$-prescription~\eqref{epterm},
by taking $\mathcal{L}_{\varepsilon}=\varepsilon \Phi_0$.
We thus find  WT identity (\ref{diag22}) in the form (see Ref.~\cite{FujPap} and~\ref{wt} for the derivation):
\bea
i \, v_0 \ = \ \lim_{\varepsilon \rightarrow 0}  \varepsilon \int \! \dr^4 y \ \! \left\lan T\lf[ \Phi^5_k(y)\, \Phi^5_k(0)\ri]\! \right\ran \, ,
\label{v_0 case}
\eea
where $k = 1,2,3$. Because the LHS differs from zero, the K\"{a}ll\'{e}n--Lehmann spectral representation of RHS implies that quantity $\varepsilon \rho(\boldsymbol{{k}}=\boldsymbol{{0}}, m_k)/m_k^2$ ($\rho$ is spectral distribution) is non-vanishing for $\varepsilon \rightarrow 0$ and thus masses {$m_k^2 \propto\varepsilon$}  ($k = 1,2,3$) due to positive definiteness of $\rho$.
This is the NG theorem, which states that the expression of $\Phi^5_k$ in the physical states representation,
called \emph{dynamical map} or \textit{Haag expansion}~\cite{Umezawa,Haag}, will contain the gapless NG fields as linear terms~\cite{Umezawa}:
\be
\Phi^5_k(x) \ = \ \sqrt{Z_{\ph^5_k}} \, \ph^5_k(x)\ + \ \ldots \, , \qquad k=1,2,3 \, .
\ee
where $Z_{\ph^5_k}$ are the wave function renormalization factors and $\ph^5_k$ are the NG fields.
Note that we have three NG fields which coincides with $\dim(G/H)$.\\

{\bf ii)} As a second case we consider the SSB pattern
\be
SU(2)_L \times SU(2)_R \times U(1)_V \ \!\longrightarrow \ \! U(1)_V \times U(1)^3_V  \, ,
\ee
which is responsible for the dynamical generation of different masses. In this case the order parameters take the form
\begin{eqnarray}
\lan \Phi_0 \ran \ = \  v_0  \ \neq \  0 \, ,  \qquad \lan \Phi_3 \ran  \ = \  v_3  \ \neq \  0  \, .
\label{diagm2}
\end{eqnarray}
The corresponding $\varepsilon$-term prescription has now the form
$\mathcal{L}_\varepsilon =  \varepsilon (\Phi_0  + \Phi_3)$. By setting $N = \{Q_1, Q_2,  \boldsymbol{Q}_5\}$ we obtain $\delta_{1(2)} \Phi_3 =  \Phi_{2} (-\Phi_1)$,
and $\delta^2_{1(2)} \Phi_3 = -\Phi_3$  and hence the WT identities (\ref{WT1}) boil down to
\bea
i\, v_3 \ = \ -\lim_{\varepsilon \rightarrow 0} \varepsilon \int \! \dr^4 y \ \! \lan T\lf[\Phi_k(y)\, \Phi_k(0)\ri] \ran \, ,  \quad k=1,2 \, .
\label{wt1}
\eea
This gives two NG modes. Moreover, Eq.~(\ref{v_0 case}) for $\Phi_{1}^5$ and $\Phi_{2}^5$ still hold, yielding another two NG fields. Finally, when  $\delta_{5,3}$
is applied to  $ \Phi_0  + \Phi_3$ we get
\bea
 i(v_3 + v_0)  \ = \  \lim_{\varepsilon \rightarrow 0 } \,  \varepsilon \int \!\!\dr^4 y \ \! \langle T[(  \Phi_3^5(y) + \Phi_0^5(y)) (\Phi_3^5(0) + \Phi_0^5(0)) ] \rangle .
\label{30.aab}
\eea
Hence, the dynamical maps of $\Phi_1,\Phi_2, \Phi_1^5, \Phi_2^5 $ and $\Phi_3^5 + \Phi_0^5$ will contain NG fields as linear terms in their Haag expansion. The number of NG fields is now five which coincides with $\dim(G/H) = \dim(G) - \dim(U(1)_V \times U(1)^3_V)$.\\

{\bf iii)} Finally, we consider the SSB scheme
\be
SU(2)_L \times SU(2)_R \times U(1)_V \longrightarrow  U(1)_V \times U(1)^3_V \longrightarrow U(1)_V \, ,
\label{SSB_scheme_3}
\ee
which is responsible for the dynamical generation of field mixing. In Appendix~\ref{ssbu1u1} we show that dynamically generation of mixing cannot occur in the breaking scheme $SU(2)_L \times SU(2)_R \times U(1)_V \rightarrow  U(1)_V \times U(1)^{\boldsymbol{m}}_V$, as it could be expected from point iii) in Section \ref{ChAl}.

Let us now introduce
\bea
\Phi_{k,m} \ = \  \overline{\boldsymbol{\psi}}\, \si_k \,  \mpsi \, ,  \qquad k=1,2,3 \, ,
\eea
where $m$ indicates that $\mpsi$ is a doublet of fields $\mpsi=\lf[\psi_1 \, \psi_2\ri]^T$ in the mass basis. 
The SSB condition in this case reads
\be \label{v1m}
\lan \Phi_{1,m} \ran \ \equiv \ v_{1,m} \ \neq \ 0 \, .
\ee
Hence we find that a \textit{necessary condition} for a dynamical generation of field mixing within chiral symmetric systems,
is the presence of  exotic  pairs in the vacuum, made up by fermions and antifermions with different masses \footnote{Generally also diagonal condensate may be present.}:
\be
\lan \overline{\psi}_i(x) \, \psi_j(x) \ran \neq \ 0 \, , \qquad i \ \neq \ j \, .
\ee
In other words, \textit{field mixing requires mixing at the level of the vacuum condensate structure}.
This conclusion is consistent with analogous results obtained in the context of QFT treatment of neutrino oscillations, in which case a \textit{flavor vacuum} has the structure of a non-trivial (Perelomov-type) condensate~\cite{BlaVit95,Mixing,BlaMavVit}. Moreover, this is an agreement with Ref.~\cite{mavroman}, where, as previously mentioned, this structure is recovered via dynamical symmetry breaking in a specific effective model generated by string-brane scattering. We remark that our result is basically model independent (the only assumption made was the global chiral symmetry), and has a non-perturbative nature. In the next section we will see that in the mean-field approximation the vacuum condensate responsible for (\ref{v1m}) formally resembles the aforementioned flavor vacuum structure.

To write down the WT identity we add the $\varepsilon$-term
\be
\mathcal{L}^{{\rm{\tiny{eff}}}}_\varepsilon(x) \ = \ \varepsilon \, \Phi_{1,m}(x) \, ,
\ee
to $\mathcal{L}^{{\rm{\tiny{eff}}}}$, which denotes the effective mass-fields Lagrangian that emerges
after the SSB $SU(2)_L \times SU(2)_R \times U(1)_V \rightarrow  U(1)_V \times U(1)^3_V$. We thus get
\bea
i\, v_{1,m} \ = \ -\lim_{\varepsilon \rightarrow 0} \varepsilon \int \! \dr^4 y \ \! \lan T\lf[\Phi_{2,m}(y)\, \Phi_{2,m}(0)\ri]  \ran \, .
\label{wtmix}
\eea
This implies a new NG field, which appears linearly in the dynamical map of $\Phi_{2,m}$. Consequently, the above two-stage SSB generates 6 NG fields.

\section{Mixing and Bogoliubov transformations}\label{meanmix}
%

It is well known~\cite{Miransky} that the manifold of degenerate ground states in the broken phase --- ordered-phase vacuum manifold,
is isomorphic to the quotient space $G/H$. We might, thus employ Perelomov group-related coherent states (CS) \cite{Perelomov} to find an explicit representation of the ensuing vacuum manifold and to carry out quantization via coherent-state functional integrals~\cite{Inomata}. The form of the interaction part of the Lagrangian  would be then reflected in the way the renormalized parameters and fields in the CS
run with the renormalization scale. This complicated model-dependent task can often be conveniently bypassed
by the mean-field approximation (MFA). In the MFA,
only quadratic operators in the Lagrangian are considered to be relevant for the description of the phase transition~\cite{Miransky,UTK}.

In the present case the symmetric-phase quasi-fields $\tilde{\psi}_1,\tilde{\psi}_2$ are massless and have a simple mode expansion
\be \label{masslessfermion}
\tilde{\psi}_j (x) \ = \ \frac{1}{\sqrt{V}}\sum_{\G k,r} \, \lf(\tilde{u}^r_{\G k,j}e^{-i |\G k|t}\tilde{\al}^r_{\G k, j}+\tilde{v}^r_{-\G k,j}e^{i |\G k|t}\tilde{\bt}^{r\dag}_{-\G k, j}\ri)e^{i \G k \cdot \G x}  \, ,
\ee
where $j=1,2$ and $\tilde{u}^r_{\G k,j},\tilde{v}^r_{-\G k,j}$ are massless spinors. In (\ref{masslessfermion}) we employed the box regularization, i.e., we enclosed our system in a box of volume $V$. Operators $\tilde{\al}^r_{\G k,j}$ and $\tilde{\bt}^r_{\G k,j}$ annihilate the
corresponding (fiducial) vacuum $|0\ran$ --- symmetric-phase mean-field vacuum. By assuming the validity of MFA we can employ
existing techniques and results involving Bogoliubov transformations~\cite{Umezawa,Miransky} to discuss the structure of vacuum manifolds in our three SSB schemes.\\

{\bf i)} As shown, e.g., in Refs.~\cite{BlaMavVit,Miransky,UTK}, the MFA vacuum for fields with dynamically generated mass
can be expressed, in terms of $|0\ran$, as~\footnote{In this Section, we denote the vacuum state for each case, with a different notation, instead of the the generic $|\Omega\ran$, used in Section~\ref{ChAl}.}
\begin{eqnarray}
|0\ran_m & = & \prod_{i=1,2}\prod_{\G k,r} \lf(\cos \te_\G k - \eta^r\sin \te_\G k \tilde{\al}^{r\dag}_{\G k,i}\tilde{\bt}^{r\dag}_{-\G k,i}\ri) |0\ran  \ = \ B(m)|0\ran \, ,
\label{mass-vacuum_1}
\end{eqnarray}
with $\te_\G k \ = \ \ha \cot^{-1}\lf({|\G k|}/{m}\ri)$ and $\eta^r=(-1)^r$. Here $m$ is the physical mass and
\be
B(m)\ = \ B_1(m) \, B_2(m) \, ,
\ee
where $B_1(m)$ and $B_2(m)$ are generators of Bogoliubov transformations, i.e.
\be
B_j(m) \ = \ \exp \lf[\sum_{\G k,r} \, \Theta_\G k \,\eta^{r}  \lf(\tilde{\al}^{r}_{\G k,j}\tilde{\bt}^{r}_{-\G k,j}-\tilde{\bt}^{r\dag}_{-\G k,j}\tilde{\al}^{r\dag}_{\G k,j}\ri)\ri] \, ,
\ee
with $j=1,2$. Above $|0\ran_m$ together with (\ref{diag1}) yield, in the large volume limit, the order parameter
\be
v_0 \ = \ 2 \, \intk \, \sin 2 \te_\G k \, .
\ee

{\bf ii)} Dynamical generation of different masses follows from a simple generalization of (\ref{mass-vacuum_1}).
In fact, it is easy to check that the  \textit{mass vacuum} has the form
\bea \label{massvac}
|0\ran_{1,2} \!\!\!\!& = &\!\!\!\!\!
\prod_{j=1,2}\prod_{\G k,r} \lf(\cos \te_{\G k, j} - \eta^r \sin \te_{\G k, j} \tilde{\al}^{r\dag}_{\G k,j}\tilde{\bt}^{r\dag}_{-\G k,j}\ri) |0\ran \ = \  B(m_1,m_2)|0\ran \, .
\eea
Here $\te_{\G k,j}  =  \ha \cot^{-1}\lf({|\G k|}/{m_j}\ri), \, j=1,2 $.  Generator $B(m_1,m_2) $
factorizes again to the product of two Bogoliubov transformations:
\be
B(m_1,m_2) \ = \ B_1(m_1) \, B_2(m_2) \, .
\ee
The ladder operators of massive fields can now be defined as (cf. e.g. Ref.~\cite{BlaMavVit})
\bea
\al^{r}_{\G k,j}  & = & \cos \te_{\G k, j} \tilde{\al}^{r}_{\G k,j} \ + \ \eta^r \sin \te_{\G k, j} \tilde{\bt}^{r\dag}_{-\G k,j}\, , \\[2mm]
\bt^{r}_{-\G k,j} & = & \cos \te_{\G k, j} \tilde{\bt}^{r}_{-\G k,j} \ - \ \eta^r \sin \te_{\G k, j} \tilde{\al}^{r\dag}_{\G k,j} \, ,
\eea
with $j=1,2$. In terms of these, we can expand the massive fields, as
\bea
\psi_j (x) & = & \frac{1}{\sqrt{V}}\sum_{\G k,r} \, \, \lf(u^r_{\G k,j}e^{-i \om_{\G k,j}t}\al^r_{\G k, j} +  v^r_{-\G k,j}e^{i \om_{\G k,j} t}\bt^{r\dag}_{-\G k, j}\ri)e^{i \G k \cdot \G x}  \, , \label{massfermion}
\eea
where $j=1,2$ and $\om_{\G k,j}=\sqrt{|\G k|^2+m^2_j}$.
The order parameters (\ref{diagm2}) are now
\bea
v_0 & = & \sum_{j=1,2}\intk \, \sin 2 \te_{\G k, j} \, , \\
v_3 & = & \intk \, \sin 2 \te_{\G k, 1}-\intk \, \sin 2 \te_{\G k, 2} \, .
\eea
\vspace{0.1mm}

{\bf iii)} The broken-phase MFA vacuum for the dynamical mixing generation has to be constructed on the mass vacuum of case ii), according to Eq.\eqref{SSB_scheme_3}. As known
(cf., e.g., Ref.~\cite{Mixing}), the relation between flavor and mass vacua can be written in the form:
\begin{eqnarray}
|0\ran_{e ,\mu} \ = \ G^{-1}_\theta(0)|0\ran_{1,2} \, ,
\label{50bbc}
\end{eqnarray}
with the generator
\be
G_\theta(t) \ = \ \exp\lf[ \intx \, \lf(\tilde{\psi}^\dag_1(x) \, \tilde{\psi}_2(x)-\tilde{\psi}^\dag_2(x) \, \tilde{\psi}_1(x) \ri)\ri] \, .
\ee
Here, we have used the sub-index ``$e,\mu$'' for the flavor vacuum in order to stay as close as possible to the usual notation used, for instance, in neutrino mixing physics, see, e.g. Ref.~\cite{BJVa}.
The flavor fields can now be written as
\bea \label{psie}
\psi_e(x) & = & \cos \theta \, \psi_1(x) \, + \, \sin \theta \, \psi_2(x)\, ,  \\[2mm]
\psi_\mu(x) & = & -\sin \theta \, \psi_1(x) \, + \, \cos \theta \, \psi_2(x) \, . \label{psimu}
\eea
The state $|0\ran_{e \mu}$, for $\theta \neq 0$ can be explicitly written as \cite{BlaVit95}:
\bea \non
|0\rangle_{e,\mu}& = & \prod_{{\bf k}}\prod_{r}\Big[(1-\sin^{2}\theta\;V_{{\bf k}}^{2})- \eta^{r}\sin\theta\;\cos\theta\; V_{{\bf k}}(\alpha^{r\dag}_{{\bf k},1}\beta^{r\dag}_{-{\bf k},2}+\alpha^{r\dag}_{{\bf k},2}\beta^{r\dag}_{-{\bf k},1})  \\[2mm]
& &+  \eta^{r}\sin^{2}\theta \;V_{{\bf k}} U_{{\bf k}}(\alpha^{r\dag}_{{\bf k},1}\beta^{r\dag}_{-{\bf k},1}-\alpha^{r\dag}_{{\bf k},2}\beta^{r\dag}_{-{\bf k},2}) +  \sin^{2}\theta \; V_{{\bf k}}^{2}\alpha^{r\dag}_{{\bf k},1}\beta^{r\dag}_{-{\bf k},2} \alpha^{r\dag}_{{\bf k},2}\beta^{r\dag}_{-{\bf k},1} \Big]|0\rangle_{1,2} \, , \label{fv}
\eea
where
\bea
&& U_\G k \ = \ \mathcal{A}_\G k  \lf(1+\frac{|\G k|^2}{\lf(\om_{\G k,1}+m_1\ri)\lf(\om_{\G k,2}+m_2\ri)}\ri)\, ,\qquad
V_\G k \ = \ \mathcal{A}_\G k \lf(\frac{|\G k|}{\om_{\G k,1}+m_1}-\frac{|\G k|}{\om_{\G k,2}+m_2}\ri)\, ,
\eea
with $\mathcal{A}_\G k =  \sqrt{\frac{\lf(\om_{\G k,1}+m_1\ri)\lf(\om_{\G k,2}+m_2\ri)}{4\om_{\G k,1}\om_{\G k,2}}}$.
This is exactly the flavor vacuum~\cite{BlaVit95,Mixing,BlaMavVit,BJSun}.

The order parameter (\ref{v1m}) assumes now the form
\bea
v_{1,m} \ = \ 2\sin 2 \theta \, \intk \, \left(\frac{m_2}{\omega_{{\bf k},2}} - \frac{m_1}{\omega_{{\bf k},1}}  \right) .
\eea
Notice that the vacuum $|0\rangle_{e,\mu}$, which is responsible for the  dynamical  generation of mixing,
contains an ``exotic'' condensate of fermion-antifermion pairs of fields with  different masses. Because of these terms, the structure of the above state is an entangled one, namely it cannot be represented in terms of product states belonging to the Hilbert spaces for $\psi_1$ and $\psi_2$. This is consistent with the  observation that flavor mixing and oscillations may equivalently be described in terms of entanglement \cite{ent}.

\smallskip

Let us remark that the above considerations have implications going beyond a purely formal level. For instance,  in the context of neutrino physics, flavor states defined as $|\nu^r_{\G k,\si}\ran \ \equiv \ \al^\dag_{\G k,\si} \, |0\ran_{e,\mu}$, lead to the 
 oscillation formula~\cite{BHV99}:
\bea
\mathcal{Q}_{\si \rightarrow \rho}(t)  & \equiv & \lan \nu^r_{\G k,\si}|Q_\rho(t)|\nu^r_{\G k,\si}\ran \non \\[2mm]
& = & \sin^2 2 \theta \, \lf(U^2_{\G k}\, \sin^2 \lf(\frac{\om_{\G k,2}-\om_{\G k,1}}{2} t\ri)+V^2_{\G k}\, \sin^2 \lf(\frac{\om_{\G k,2}+\om_{\G k,1}}{2}t\ri)\ri) \, , \qquad \si \neq \rho =e,\mu \, ,
\label{55ab}
\eea
where
\be
Q_\rho(t) \ \equiv \ \intx \, \psi^\dag_\rho(x) \, \psi_\rho(x) \, ,
\ee
are the ``physical" (non-conserved) flavor charges defined in terms of the flavor fields (see Eqs.\eqref{psie}, \eqref{psimu}). Eq.~(\ref{55ab}) exhibits phenomenological corrections with respect to the usual Pontecorvo formula~\cite{Pontecorvo}, which is recovered only in the ultra-relativistic limit.  
\section{Conclusions and Discussion}\label{conclusioni}

In this paper we have discussed three types of dynamical symmetry-breaking schemes for a generic Lagrangian density with $G=SU(2)_L \times SU(2)_R \times U(1)_V$ chiral symmetry: i)  $G \rightarrow U(2)_V$, ii) $G \rightarrow U(1)_V \times U(1)^3_V$ and iii) the two-step SSB $G \rightarrow U(1)_V \times U(1)^3_V \rightarrow U(1)_V$. We demonstrated that these symmetry-breaking schemes lead to: i) a single dynamically generated  mass for the doublet field, ii) two different dynamically generated  masses for the fields (without mixing) and iii) dynamically generated mixing among the two fields.

Our analysis was based on an algebraic and hence manifestly non-perturbative point of view.
In particular, we employed Umezawa's $\varepsilon$-term prescription alongside with WT identities for SSB to gain information
about NG bosons and ensuing set of ground states. The explicit form of the ground states that are responsible for dynamical generation of masses and field mixing was obtained in MFA where they were phrased in terms of generators of
Bogoliubov transformations acting on fiducial vacuum states.

Our key finding is that the vacuum state, responsible for dynamical mixing generation, exhibits similar condensate structure as the flavor vacuum defined in the context of neutrino
oscillations~\cite{BlaVit95,Mixing,BlaMavVit,BJSun}.
Hence, in order to generate field mixing dynamically, a non-trivial vacuum structure is required. This was reflected in Section~IV via appearance of a Bogoliubov transformation. Let us note that the condition $\lan \boldsymbol{\Phi}^5\ran =0$ does not affect considerably our main results. Indeed, if $\lan \boldsymbol{\Phi}^5\ran \neq 0$ we would still need mixed vacuum condensates, as it can be verified by means of table in Appendix~\ref{tabu1} and Eq.~\eqref{cssb}. The only difference would be  that  expressions such as Eq.~\eqref{fv} will acquire extra phase factors.

Let us finally add some comments. A common feature of SSB is the appearance of topological defects~\cite{Umezawa}. The number of such defects is related to the quench time of SSB in which they are formed via the Kibble--Zurek mechanism~\cite{KibZur}. On the other hand, type of defects in 3-D configuration space is determined by a non-trivial homotopy group $\pi_n(G/H)$ ($n= 0,1,2$).
By analogy with condensed matter systems we might expect that defects formed might provide an important observational
handle on the dynamics of the mixing-related SSB transition.

It is known~\cite{maguejo}  that Lorentz symmetry may be spontaneously broken by the flavor vacuum, in the sense that the corresponding dispersion relations of states constructed as Fock excitations of the flavor vacuum are modified as compared to the standard Lorentz covariant ones. In this sense, one can discuss flavor mixing in a fixed frame, such as finite temperature situations, which break Lorentz symmetry.

%

\section*{Acknowledgements}
It is pleasure to acknowledge helpful conversations with
G.~Vitiello. P.J.  was  supported  by the Czech  Science  Foundation Grant No. 17-33812L.
The research of N.E.M. was supported partly by the STFC Grant ST/L000258/1.
 NEM acknowledges the hospitality of IFIC Valencia through a
Scientific Associateship ({\it Doctor Vinculado}).
\appendix
\section{Ward--Takahashi identities and Umezawa's $\varepsilon$-term prescription}\label{wt}

Here we briefly review the proof of Ward--Takahashi identities with the Umezawa's $\varepsilon$-term prescription~\cite{Umezawa}.
To this end, we consider the $n$-point Green's function (possible internal
indexes are suppressed)
\be
G^n(x_1,x_2, \ldots, x_n)\ = \ \lan \mathrm{T}[\phi(x_1)\, \phi(x_2) \,\ldots \, \phi(x_n)] \ran \, .
\ee
By employing the fact that $\phi$ transforms under the influence of group generators $N_1, N_2, \ldots N_n$ as
\be
 \phi'(x)  =  \phi(x) \, + \, \de \phi(x) \, , \;\;\;\; \de \phi(x) \ = \ \sum_{k=1}^n \ep_k \de_k \phi(x) \, , \label{groupsy}
\ee
with
\be
\de_k \phi(x) \ \equiv \ i \, [N_k,\phi(x)] \, ,
\ee
one can show that~\cite{Umezawa}
\bea\non
\frac{\pa}{\pa t}\lan \mathrm{T}[N(t) \, \phi(x_1) \, \phi(x_2) \, \ldots \, \phi(x_n)] \ran   &=& \ \sum^n_{j=1} \delta(t-t_j)\lan \mathrm{T}[\phi(x_1) \, \ldots  \, \lf[N(t) \, , \, \phi(x_j)\ri]\ldots \phi(x_n)]\ran \non \\[2mm]
&&+ \ \lan \mathrm{T}\lf[\dot{N}(t) \, \phi(x_1) \, \phi(x_2) \, \ldots \, \phi(x_n)\ri] \ran \, , \label{dergre}
\eea
with
\be
N(t)\ \equiv \ \sum_{k=1}^n\ep_k \, N_k(t) \, . \label{ridgen}
\ee
By employing Noether's theorem in the form~\cite{Miransky, Umezawa}
\be
\dot{N}(t)\ = \ \intx \, \de \mathcal{L}(x) \, .
\label{A.5aa}
\ee
we can write
\bea \non
\frac{\pa}{\pa t}\lan \mathrm{T}[N(t) \, \phi (x_1) \, \phi(x_2) \, \ldots \, \phi(x_n)] \ran
& =& \ -i\sum^n_{j=1} \delta(t-t_j)\lan \mathrm{T}[\phi(x_1) \, \ldots  \, \de \phi(x_j) \, \ldots \, \phi(x_n)]\ran \non \\
&& + \ \intx \, \lan \mathrm{T}[\de \mathcal{L}(x) \, \phi(x_1) \, \phi(x_2) \, \ldots \, \phi(x_n)] \ran \, . \label{dergre2}
\eea
Let us now integrate both sides of (\ref{dergre2}) for $t \in (-\infty, \infty)$. This gives	
\bea
&&i\sum^n_{j=1}\lan \mathrm{T}[\phi(x_1) \, \ldots \, \de\phi(x_j) \, \ldots \, \phi(x_n)] \ran  = \ \int \!\!\mathrm{d}^4 x \, \lan \mathrm{T}[\de \mathcal{L}(x)\,\phi(x_1) \, \phi(x_2) \, ,  \ldots \phi(x_n)] \ran \, , \label{WT1}
\eea
which is the form of the \textit{Ward--Takahashi identity} employed in the main text.
Note that the integral of the LHS of Eq.~\eqref{dergre2} could not vanish
if in the spectrum would be present infinite range correlations, as those due to the NG modes in SSB.
To avoid this technical difficulty we employ Umezawa's  $\varepsilon$-term prescription~\cite{Umezawa,FujPap},
i.e., we add to $\mathcal{L}$ an explicit breaking term
\be \label{epterm}
\mathcal{L}_\varepsilon(x) \ = \ \varepsilon \, \Phi(x) \, .
\ee
where $\Phi$ is an order-parameter operator characterizing  a particular SSB scheme.
At the end of calculations the limit $\varepsilon \rightarrow 0$ has to be taken, after the thermodynamical limit.

As an illustration we now sketch the derivation of Eq.~\eqref{v_0 case}. The other relations follow in a similar way.
If we take $n=1$:
\begin{eqnarray} \!\!
i\langle \delta \phi (0) \rangle \ = \ \int \!\! d^4 y \,\langle T [\delta \mathcal{L}(y) \, \phi(0)]\rangle\, ,
\label{diag22}
\end{eqnarray}
In particular we will choose $\phi=\de \Phi $, so that:
\begin{eqnarray} \label{wtd}
i \lan \delta^2 \Phi(0) \ran \ = \ \lim_{\varepsilon \rightarrow 0} \varepsilon \, \int \! \dr^4 y \ \! \lan T\lf[\delta \Phi(y) \delta \Phi(0)\ri] \ran\, .
\end{eqnarray}
If we take ${{N}}_k = {{Q}}_{5,k}$, $\Phi =  \Phi_0$ and $\phi = \delta_{5,k} \Phi_0$, with $k=1,2,3$ we get
\bea
i \lan \de^2_{5,k} \Phi_0\ran \ = \ \lim_{\varepsilon \rightarrow 0}  \varepsilon \int \! \dr^4 y \ \! \lan T\lf[ \delta_{5,k}\Phi_0(y)\, \delta_{5,k}\Phi_0(0)\ri]  \ran \, .
\label{v_0almost}
\eea
Because $\delta_{5,k} \Phi_0 = -i\Phi^5_k$ and $\delta^2_{5,k} \Phi_0 = -\Phi_0$, we finally arrive at Eq.~\eqref{v_0 case}.

\section{Note on SSB scheme $SU(2)_L \times SU(2)_R \times U(1)_V  \rightarrow U(1)_V \times U(1)^\boldsymbol{m}_V$}\label{ssbu1u1}
Let consider the SSB scheme \
\be
SU(2)_L \times SU(2)_R \times U(1)_V \longrightarrow  U(1)_V \times U(1)^\boldsymbol{m}_V \, ,
\ee
which would correspond to mixing generation according to classical reasonings presented in Section \ref{ChAl}. This SSB scheme is characterized by the order parameters
\bea
\lan \Phi_k \ran  =  v_k  \neq  0 \, , \qquad k\ = \ 0,1,2,3  \, .
\eea
The $\varepsilon$-term prescription assumes now the form
$\mathcal{L}_\varepsilon =  \varepsilon \sum^3_{k=0}\Phi_k $. It is easy to check that  Eqs.~\eqref{v_0 case}, \eqref{wt1} and~(\ref{30.aab})
 have to be replaced by
\begin{eqnarray} \non
 i(v_0 + v_k) & = &   \lim_{\varepsilon \rightarrow 0 } \,  \varepsilon \, \int \!\!\dr^4 y \ \! \lan T[(\Phi_k^5(y) + \Phi_0^5(y))(\Phi_k^5(0) + \Phi_0^5(0))]\rangle \, , \\[2mm] \nonumber
 - i (v_2 + v_j) & = & \lim_{\varepsilon \rightarrow 0} \, \varepsilon \, \int \!\! \dr^4 y \, \langle T[(\Phi_2(y)-\Phi_j(y))(\Phi_2(0)-\Phi_j(0))] \rangle \, ,  \\[2mm]
- i (v_1 + v_3) & = & \lim_{\varepsilon \rightarrow 0} \, \varepsilon \, \int \!\! \dr^4 y \, \langle T[(\Phi_1(y)-\Phi_3(y))(\Phi_1(0)-\Phi_3(0))] \rangle \, ,
\end{eqnarray}
with $k=1,2,3$ and $j=1,3$. Consequently, the NG modes will be associated with following
fields: $\Phi_2-\Phi_3$, $\Phi_{1}-\Phi_{3}$, $\Phi_{2}^5+\Phi^5_0$, $\Phi_{1}^5 + \Phi_{0}^5$ and $\Phi_{3}^5+ \Phi_{0}^5$.
The number of NG modes is thus $5$  which coincides with $\dim(G/H)$. Evidently this SSB pattern cannot describe dynamical mixing generation because it is equivalent to the case ii) of Section \ref{LBcon}.\\

\section{Table of first variations} \label{tab}

Below we list the first variations of composite operators introduced in Section \ref{LBcon} which are used in the main text:

\begin{table}[h]
\label{tabu1}
\begin{tabular}{|c|c|c|c|c|c|c|c|c|}
\hline \hline
 $$ & $ \hat{\Phi}_0 $ & $ \hat{\Phi}_1 $ & $ \hat{\Phi}_2 $ &  $\hat{\Phi}_3$ & $ \hat{\Phi}^5_0 $ & $ \hat{\Phi}^5_1 $ & $ \hat{\Phi}^5_2 $ & $\hat{\Phi}^5_3$\\
\hline
 $\de_0$ & $ 0 $ & $ 0 $ & $0$ & $0$ & $ 0 $ & $ 0 $ & $ 0 $ & $ 0 $\\
\hline
  $\de_1$ & $ 0 $ & $0$ & $-\hat{\Phi}_3$ & $\hat{\Phi}_2$ & $ 0 $ & $ 0 $ & $-\hat{\Phi}_3^5$ & $ \hat{\Phi}_2^5 $\\
\hline
  $\de_2$ & $ 0 $ & $\hat{\Phi}_3$ & $0$ & $ -\hat{\Phi}_1 $ & $ 0 $ & $ \hat{\Phi}^5_3 $ & $0$ & $ -\hat{\Phi}^5_1$\\
\hline
  $\de_3$ & $ 0 $ & $-\hat{\Phi}_2$ & $\hat{\Phi}_1$ & $0 $  & $ 0 $ & $ -\hat{\Phi}^5_2 $ & $\hat{\Phi}^5_1$ & $0 $ \\
\hline
 $\de_{5,0}$ & $ -i \hat{\Phi}^5_0$ & $ -i \hat{\Phi}^5_1$ & $-i\hat{\Phi}_2^5$ & $ -i \hat{\Phi}^5_3 $ & $ -i \hat{\Phi}_0 $ & $ -i \hat{\Phi}_1$ & $-i\hat{\Phi}_2$ & $ -i \hat{\Phi}_3 $\\
\hline
  $\de_{5,1}$ & $ -i \hat{\Phi}^5_1 $ & $-i \hat{\Phi}^5_0$ & $0$ & $ 0 $ & $ -i \hat{\Phi}_1 $ & $-i \hat{\Phi}_0$ & $0$ & $ 0 $\\
\hline
  $\de_{5,2}$ & $ -i\hat{\Phi}^5_2 $ & $0$ & $-i\hat{\Phi}_0^5$ & $0$ & $-i\hat{\Phi}_2$ & $ 0 $ & $-i\hat{\Phi}_0$ & $ 0 $\\
\hline
  $\de_{5,3}$ & $ -i\hat{\Phi}^5_3 $ & $0$ & $0$ & $-i \hat{\Phi}^5_0 $ & $ -i\hat{\Phi}_3 $ & $ 0 $ & $0$ & $-i \hat{\Phi}_0 $\\
\hline \hline
\end{tabular}
\end{table}

\pagebreak

\end{document}